\documentstyle[12pt]{article}

\newcommand{\mysection}[1]{\setcounter{equation}{0}\section{#1}}

\newcommand{\nc}{\newcommand}
\nc{\beq}{\begin{equation}} \nc{\eeq}{\end{equation}}
\nc{\beqa}{\begin{eqnarray}} \nc{\eeqa}{\end{eqnarray}}
\nc{\lsim}{\begin{array}{c}\,\sim\vspace{-21pt}\\< \end{array}}
\nc{\gsim}{\begin{array}{c}\sim\vspace{-21pt}\\> \end{array}}

\begin{document}

\begin{titlepage}
\begin{center}
{\hbox to\hsize{May 1994 \hfill JHU-TIPAC-940005}}
{\hbox to\hsize{hep-ph/9405345 \hfill MIT-CTP-2309}}
{\hbox to\hsize{ \hfill NSF-ITP-94-48}}

\bigskip

\bigskip

{\Large \bf
The $R$ Axion From Dynamical Supersymmetry Breaking} \\

\bigskip

{\bf Jonathan Bagger,\footnotemark[1]
Erich Poppitz\footnotemark[1]$^,$\footnotemark[2]}\\

\smallskip

{\small \it Department of Physics and Astronomy

The Johns Hopkins University

Baltimore, MD 21218, USA }

\smallskip
{\small and}
\smallskip

{\bf Lisa Randall}\footnotemark[3]\\

\smallskip

{ \small \it Center for Theoretical Physics

Laboratory for Nuclear Science and Department of Physics

Massachusetts Institute of Technology

Cambridge, MA 02139, USA

and

The Institute for Theoretical Physics

Santa Barbara, CA 93106, USA }

\bigskip

{\bf Abstract}\\[-0.05in]
\end{center}
{\small All generic, calculable models of dynamical supersymmetry breaking
have a spontaneously broken $R$ symmetry  and therefore contain an $R$ axion.
We show that the axion is
massive in any model in which the cosmological constant is fine-tuned
to zero through an explicit $R$-symmetry-breaking constant. In visible-sector
models, the axion mass is in the 100 MeV range and
thus evades astrophysical bounds. In nonrenormalizable hidden-sector models,
the mass is of order of the weak scale and can have dangerous cosmological
consequences similar
to those already present from other fields.
In renormalizable hidden-sector models, the axion
mass is generally quite large, of order $10^7$ GeV. Typically, these axions
are cosmologically safe. However, if the dominant decay mode is to gravitinos,
the potentially large gravitino abundance that arises from axion decay after
inflation might affect the successful predictions of big-bang nucleosynthesis.
We show that the upper bound on the reheat temperature after standard inflation
can be competitive with or stronger than bounds from thermal gravitino
production,
depending on the model and the gravitino mass.}

\bigskip

\bigskip

\footnotetext[1]{Supported in part by
NSF grants PHY90-96198 and PHY89-04035, and by
the Texas National Research Laboratory Commission
under grant RGFY93-292.}
\footnotetext[2]{After August 1, 1994:
{\it The Enrico Fermi Institute,
University of Chicago, 5640 Ellis Avenue, Chicago, IL 60637.}}
\footnotetext[3]{NSF Young Investigator Award,
Alfred P.~Sloan
Foundation Fellowship, DOE Outstanding Junior
Investigator Award. Supported in part
by DOE contract DE-AC02-76ER03069, by NSF grant PHY89-04035, and by
the Texas National Research Laboratory Commission
under grant RGFY92C6.}

\end{titlepage}

\renewcommand{\thepage}{\arabic{page}}
\setcounter{page}{1}

\baselineskip=18pt

\mysection{Introduction}

The primary attraction of supersymmetry is that it suggests
a solution to the gauge hierarchy problem.  However, it does not
necessarily explain the ratio $m_W/M_P$.  Dynamical models
of supersymmetry breaking offer a possible reason why $m_W/M_P
\simeq 10^{-17}$.  In such models all scales arise
from the Planck scale through dimensional transmutation
\cite{witten}.  Supersymmetry is unbroken to all orders of perturbation
theory, but nonperturbative effects, proportional to $e^{- O(1) 8\pi^2/g^2}$,
can generate a ground state that breaks supersymmetry \cite{ADS3}.

At present there is only a limited class of known models where dynamical
supersymmetry breaking is realized in a calculable way.
Some criteria for supersymmetry breaking have been
proposed \cite{ADS1,ADS2}; but it is not
yet clear whether this exhausts the class of viable models.
Reference \cite{NS} considered the class of
generic, calculable models, where
all terms consistent with the symmetries
are present, and the low-energy effective lagrangian does not contain
strongly-interacting gauge
fields. It was shown that for
such models a spontaneously broken $R$ symmetry is necessary and sufficient
for dynamical supersymmetry breaking. This spontaneously broken symmetry
implies the existence of a Goldstone boson, called the $R$ axion.

In this paper we study the properties of the $R$ axion.
We show that any generic calculable model of
supersymmetry breaking always has a trivial analog for which
the axion is massive, namely the same model with a constant
term added to the superpotential. Although the model is
technically nongeneric because
it includes a single term that breaks $R$ symmetry, it
is still a reasonable candidate for a supersymmetry-breaking
sector. In fact, this constant is necessary in any realistic model to
obtain a vanishing cosmological constant.

We also consider cosmological and astrophysical constraints on the
$R$ axion.  When supersymmetry is broken in a visible sector,
the axion is sufficiently heavy to avoid astrophysical problems.
This is important because it implies that the existence of an $R$ axion
is not sufficient cause to dismiss these models.

When supersymmetry is broken in a nonrenormalizable hidden sector,
the axion mass is generally of order the weak scale.  Such an
axion can have dangerous cosmological consequences. However,
it is already known that models of this sort generally contain
singlets with similar problems \cite{CFKRR,bkn}. One
might hope that all the difficulties will be solved
by the same mechanism. (See, for example, \cite{rt}.)

Most of this paper focuses on the axion in renormalizable
hidden-sector models.  The axion in these models
is quite heavy, with mass of order $10^7 {\rm GeV}$.  We point
out that this axion can be a new source of gravitinos and that this
leads to a new constraint on the reheat temperature of the universe after
inflation. Our conclusion is that it is not necessary to eliminate
the axion to build a successful model of dynamical supersymmetry breaking.

As we will see, in all models the $R$-axion mass arises because the
cosmological constant
must be canceled in any successful theory of dynamical supersymmetry
breaking. We do not in any way attempt to solve the cosmological constant
problem; our point is that until one understands better its solution,
one cannot dismiss models that contain an axion. This is of particular
relevance to visible-sector models.

In section 2 of this paper we discuss the relation between $R$ symmetry
and the cosmological constant. We explain the origin of the axion
mass, and consider three types of models: visible-sector models,
nonrenormalizable hidden-sector models, and renormalizable hidden-sector
models. In section 3 we consider renormalizable hidden-sector
models in more detail. We study the simplest such model, based on
a SU(3) $\times$ SU(2) gauge group, which we call the 3-2 model. Because
supersymmetry breaking
occurs in the weak-coupling regime, the spectrum and its properties
can be calculated in a controlled expansion. This model provides
a useful template for renormalizable hidden-sector models.
In section 4 we study the relevant cosmological constraints on
renormalizable hidden-sector
models. We find that if the reheat temperature of the universe after
inflation is too high, a large number of gravitinos may be produced by
coherent oscillations of the $R$ axion field. Their subsequent decay may lead
to dissociation of the light elements,
in conflict with big-bang nucleosynthesis.
We compare this bound on the reheat temperature with the bound from
thermal production of gravitino and find it competitive (for smaller
gravitino mass) or stronger (for larger gravitino mass). In section
5 we summarize our results.

\mysection{The Source of Axion Mass}

In this section, we discuss the axion mass in three scenarios
for dynamical supersymmetry breaking \cite{bkn}: nonrenormalizable
hidden-sector models (NRHS), renormalizable hidden-sector models (RHS),
and visible-sector models. In the first class of models, supersymmetry
is broken only when supergravity couplings are included (that is, supersymmetry
is unbroken in the $M_P \to \infty$ limit). In the second class, supersymmetry
is broken in the flat-space limit, and supersymmetry breaking is communicated
to the visible world through Planck-mass-suppressed interactions associated
with supergravity. In the final class of models, both supersymmetry breaking
and the communication of supersymmetry breaking are achieved through
renormalizable couplings.

In supergravity theories, the tree-level scalar potential takes the
following form
\cite{WB}:
\begin{eqnarray}
\label{v}
V&=&V_D~+~V_F~, \\[2mm]
\label{vd}
V_D&=&{1\over 2}~g^2~D^a D^a ~,\\
\label{vf}
V_F&=&\exp(K/M_P^2)~\left(\left[W_i + {K_i\over M_P^2} W\right]
K^{-1}_{i j*}\left[W^*_{j*} + {K_{j*}\over M_P^2} W^*\right] ~-~
{3 W W^* \over M_P^2}\right)~,
\end{eqnarray}
where $W$ and $K$ are the superpotential and the K\"ahler potential,
respectively, the $D^a$ are the $D$ terms of the various gauge groups,
and we have defined $M_P$ to absorb a factor of $\sqrt{8\pi}$.
Supersymmetry is spontaneously broken if $\langle \exp(K/2 M_P^2)[W_i
+ (K_i/ M_P^2) W ]\rangle$ is nonzero.\footnote{We ignore the possibility
of a Fayet-Iliopoulos $D$-term.}

{}From these expressions, we see that supersymmetry can be
spontaneously broken, with no cosmological constant, if the
superpotential contains a constant $W_0$ that is adjusted
to cancel the vacuum energy.  If there are no Planck-scale
vevs, this implies
\beq
\label{W0MS}
W_0 ~=~ {1\over \sqrt{3}}~ M^2_S M_P
\eeq
to leading order in $1/M_P$, where $M^4_S =
\langle W_i K^{-1}_{ij*} W^*_{j*} \rangle $ denotes
the scale of supersymmetry breaking.

If the original superpotential $W_2$ preserves an $R$ symmetry, it carries
$R$-charge 2, whereas any constant term has $R$-charge zero. Therefore the
cosmological term $W_0$ explicitly breaks the $R$ symmetry, which implies
that the $R$ axion is a
massive pseudo-Goldstone boson. The mass term arises from the cross terms
between the $R$-symmetry-breaking constant and the $R$-preserving terms
in the superpotential. The general formula for the axion mass follows from
(\ref{vf}), with
superpotential $W = W_2 + W_0$,
\beq
\label{Generalaxionmass}
m_a^2 ~= ~ {8\over f_a^2}~{W_0~\big\vert \langle
W_{2~i}~K_{ij*}^{-1}~K_{j*}~-~3 W_2 \rangle
\big\vert \over M_P^2}~.
\eeq
Here $f_a$ is the axion coupling. (See section 3.)
\beq
f_a^2 ~=~  2 ~r_i~r_j~v_i~v_j^*~\langle K_{ij*}\rangle ~,
\eeq
where $r_i$ and $v_j$ are the $R$ charges and vevs of the fields, respectively.
Note that this mass arises from an $F$ term, unlike the soft scalar masses of
hidden-sector models.

This explicit violation of $R$ symmetry does not contradict the
theorem of Nelson and Seiberg \cite{NS}, since the model is now technically
nongeneric because we have not added all terms that violate $R$ symmetry.
However, it is not inconceivable that the physics responsible for solving
the cosmological
constant problem communicates only through gravity, and respects different
symmetries than the supersymmetry-breaking sector of the theory.
Furthermore, the proof that an $R$ symmetry is both necessary and sufficient
relies on possible flat directions in the potential \cite{NS}. The constant
term is irrelevant to the argument, so any generic theory has a trivial
nongeneric counterpart for which the cosmological constant vanishes and the
axion is massive.

For models with dynamical supersymmetry breaking, the general
formula (\ref{Generalaxionmass}) for the axion mass becomes
\beq
\label{generalaxionmass}
m_a^2~\simeq~{1\over f^2}~{M_S^2~\Lambda^3\over M_P}~,
\eeq
where $\Lambda$ is related to the scale of the strong dynamics.
Equation (\ref{generalaxionmass}) applies for all the
models we will discuss.

In renormalizable hidden-sector models, the scale of supersymmetry
breaking $M_S \simeq f \simeq \Lambda$, so $m_a^2 \simeq M^3_S/M_P$.
In such models $M^2_S \simeq m_W M_P$, so $m_a \simeq \sqrt{m_W M_S}$,
or about $10^7$ GeV.
This mass is enhanced over the
other soft masses by a factor of $(M_P/m_W)^{1/4}$.
Because of the large mass, the axion decays quickly.
It is cosmologically safe, except
when it decays predominantly to stable
heavy particles, as we will discuss.

In nonrenormalizable hidden-sector models, the energy density at the
minimum of the potential is of order $M_S^4~\simeq ~\Lambda^6/M_P^2$,
and $f \simeq M_P$.  The factor of $1/M_P^2$ follows from the fact that
supersymmetry is broken by nonrenormalizable Planck-mass-suppressed
terms. Therefore, in NRHS models, the axion mass is of order
$m_a \simeq M^2_S/M_P \simeq m_W$, where we have again used the
fact that $M^2_S \simeq m_W M_P$.  This implies
that the axion in NRHS models suffers from
the same cosmological problems that have been identified for
moduli fields \cite{CFKRR,bkn}, or any other fields that have flat potentials
up to nonrenormalizable terms, and whose potential is determined by the
supersymmetry-breaking sector. We do not address the
cosmological problems of such fields in this paper. (See, however,
ref. \cite{rt}.)

Finally, the axion of visible-sector models is also massive. In such
models, supersymmetry breaking is typically at a much lower scale
because it is communicated to the visible world through gauge
interactions. Therefore the axion mass is much smaller, as can be
seen from (\ref{generalaxionmass}).
If the axion can be produced in a supernova,
its mass must exceed 10 MeV so that the supernova does not cool too
quickly \cite{raffelt}.
For visible-sector models with $M_S$ greater than $10^5$
GeV, there is no problem. For lower values of $M_S$,
an alternative source of axion mass may be required. (For example,
there is a contribution to the axion mass if the associated U(1) is anomalous
with respect to a
gauged symmetry \cite{dn}.) However, visible sectors seem to require
a symmetry-breaking scale as high as $10^6$ GeV because masses arise
through multiloop graphs \cite{dn}. With such a large scale of
supersymmetry breaking, the axion is sufficiently heavy to be in accord with
astrophysical bounds.

\mysection{An Explicit Calculation of the Axion Mass:\newline
the 3-2 Model}

\subsection{The Model}

The simplest known calculable model with dynamical supersymmetry breaking
is based on two-flavor supersymmetric SU(3) QCD with
gauged SU$(2)_L$ flavor symmetry \cite{ADS1,ADS2}. The model is remarkable
because the
supersymmetry-breaking ground state and the low-energy particle spectrum
can be found in a controlled weak-coupling approximation.

To describe the model, we denote the left and right quark chiral superfields
by $Q$ and $\bar{Q}$. Under the SU(3) $\times$ SU(2) gauge symmetry, they
transform as follows ($\bar{Q}^i_{~\alpha} ~ \equiv ~(\bar{D}^i, \bar{U}^i )$):
\begin{eqnarray}
\label{Q}
Q^{~\alpha}_i & \sim & (3,2)\; , \nonumber \\
\bar{U}^i & \sim & (\bar{3},1)\; , \\
\bar{D}^i & \sim &(\bar{3},1)\; ,\nonumber
\end{eqnarray}
where Greek and Roman letters denote SU(2) and SU(3) indices,
respectively. Cancellation of the Witten anomaly requires another SU(2)
doublet,
\beq
\label{L}
L^{\alpha} ~\sim ~(1, 2) \;.
\eeq
The particle content of the model is similar to that of the minimal
supersymmetric
standard model without the right-handed electron and Higgs superfields.

Apart from the gauge symmetries, the 3-2 model has two anomaly-free global
symmetries:
U$(1)_Y$ hypercharge and U$(1)_R$. The hypercharge assignments are
like those in the minimal supersymmetric standard model:
\begin{eqnarray}
\label{Ycharges}
Y(Q) &=& 1/6~,\nonumber \\
Y(\bar{U}) &=& -2/3~,\nonumber \\
Y(\bar{D}) &=& 1/3~, \\
Y(L) &=& -1/2 ~.\nonumber
\end{eqnarray}
Under the nonanomalous $R$ symmetry the charges of the matter
superfields
are given by:\footnote{Recall that the $R$ charge of a fermion
in a chiral multiplet $\Phi$ of charge $R_{\Phi}$ equals $R_{\Phi}-1$
\cite{WB}.}
\begin{eqnarray}
\label{Rcharges}
R(Q) &=& 1~,\nonumber \\
R(\bar{U}) &=&R(\bar{D})~=~ 0~,\\
R(L) &=& -3 ~.\nonumber
\end{eqnarray}
The gauginos carry $R$ charge $-1$.

The K\"ahler potential of the model takes the usual form for a renormalizable,
supersymmetric theory
\beq
\label{kahler}
K ~=~ Q^{\dagger}~Q
{}~+ ~\bar{Q}~\bar{Q}^{\dagger}~ +~ L^{\dagger}~L \; .
\eeq
In (\ref{kahler}) the SU(2) and SU(3) gauge superfields are not
written, but are assumed to be coupled in the usual way \cite{WB}.

In the absence of a superpotential, the scalar potential vanishes
for a number of flat directions in field space, and the ground state
is undetermined at the classical level \cite{ADS1}. The equations that
determine
the flat directions are
\begin{equation}
\label{flatdirectionsSU3}
Q^{\dagger ~m}_{\alpha}~Q_{l}^{~\alpha}~ - ~\bar{Q}^{m}_{~\alpha}
{}~\bar{Q}^{\dagger ~\alpha}_{l}~ =~ 0
\end{equation}
for the SU(3) $D$-terms, and
\begin{equation}
\label{flatdirectionsSU2}
Q^{\dagger ~i}_{\alpha}~Q_{i}^{~\beta} ~+~
L^{\dagger}_{\alpha}~L^{\beta}~ =~ {1\over2}~\delta^{~\beta}_{\alpha}
{}~( Q^{\dagger}~Q ~+~ L^{\dagger}~L )
\end{equation}
for the SU(2) $D$-terms. Up to local symmetries, the solutions to
these equations are parametrized by six real variables.

Let us consider the theory expanded around a solution of
(\ref{flatdirectionsSU3}), (\ref{flatdirectionsSU2}), such that the
scale $v$ of the vacuum expectation values of the scalar fields obeys
\beq
\label{scale}
v ~\gg \Lambda_3 \; ,
\eeq
where
\beq
\label{Lambda3}
\Lambda_3 ~=~ v~ {\rm exp}\left(-{ 8\pi^2\over g_3(v)^2~b_0}\right)
\eeq
is the scale where the SU(3) gauge coupling $g_3$ becomes
strong\footnote{For simplicity, we assume that the SU(2) gauge coupling
$g_2 ~\ll~g_3$.} and $b_0$ is the one-loop coefficient of the beta
function. For such vacua, the theory is in the weak-coupling regime. This
vacuum completely breaks the SU(2) and SU(3) gauge
symmetries, so the vector supermultiplets are massive. Supersymmetry
is unbroken along the flat directions, and 11 out of the 14 matter
chiral superfields are massive as well. The remaining 3 chiral
superfields are massless.

At energies below the scale $\Lambda_3$, the low-energy effective theory
can be described in terms of the following three gauge-invariant
chiral superfields
\begin{eqnarray}
\label{lightfields}
X_1 &=& Q\; \bar{D}\; L\;,\nonumber \\
X_2 &=& Q\; \bar{U}\; L\;,\;\\
X_3 &=& {\rm det}~ \bar{Q}_{\alpha}\;Q^{\beta} \; .\nonumber
\end{eqnarray}
Their scalar components parametrize the six flat directions
(\ref{flatdirectionsSU3}), (\ref{flatdirectionsSU2}) of the potential.

The superpotential is such that a global U$(1)_Y ~\times$ U$(1)_R$ symmetry
is preserved:
\beq
\label{superpotential}
W ~=~ \lambda \; X_1\;+\; 2 \; {\Lambda_3^7 \over
X_3}\;.
\eeq
The first term is the usual renormalizable superpotential. The second is
generated by nonperturbative effects. Its coefficient can be calculated
in the weak coupling expansion around a constrained instanton in a vacuum
that obeys (\ref{scale}) \cite{ADS1}.

When $\lambda = 0$, the scalar potential of the model (\ref{kahler},
\ref{superpotential}) does not have a minimum at a finite value of the
fields, so the theory does not have a ground state \cite{ADS1}. For the
case when
\beq
\label{lambda}
\lambda ~\ll~g_2~\ll~g_3~\ll 1 \; ,
\eeq
the scalar potential has almost-flat directions, given by the solutions to
(\ref{flatdirectionsSU3}), (\ref{flatdirectionsSU2}). The potential now has a
minimum at finite values $v$ for the fields,
of order
\beq
\label{vevs}
v ~\simeq ~{\Lambda_3\over\lambda^{1/7}}\; .~
\eeq
This value is such that the weak coupling assumption (\ref{scale}) is
self-consistent, so the theory can be analyzed perturbatively.

At this minimum, the vacuum energy is nonzero and supersymmetry is
spontaneously
broken\footnote{In fact, supersymmetry is broken even when the theory is
strongly-coupled.}.
In this paper we restrict our attention to the weak-coupling regime, where the
spectrum can be computed using the effective low-energy theory along the
almost-flat directions. This procedure is described in the next section.

\subsection{The Low-Energy Sigma Model and its Spectrum}

In this section we derive the gauge-invariant low-energy effective field
theory, valid below the scale $\Lambda_3$. We find the spectrum of all
particles lighter than this scale.

In the limit (\ref{lambda}), the superpotential can be treated as
a perturbation on the theory without a superpotential. Therefore
the K\"ahler potential of the effective theory is given by the projection
of (\ref{kahler}) onto the fields $X_1, X_2$ and $X_3$
(\ref{lightfields}) that span the flat directions of the potential. The
low-energy theory is a supersymmetric nonlinear sigma model with
coordinates $X_1, X_2$ and $X_3$.

{}From the equations for the SU(3) flat directions (\ref{flatdirectionsSU3}),
it follows that
\begin{equation}
\label{projection1}
Q^{\dagger}~Q
{}~=~ \bar{Q}~\bar{Q}^{\dagger}
\end{equation}
along the flat directions. Using the definitions of the light fields
(\ref{lightfields}) and the equations for the flat directions
(\ref{flatdirectionsSU3}), (\ref{flatdirectionsSU2}), we find
\begin{eqnarray}
X^{\dagger}_1~ X_1~ + ~X^{\dagger}_2~ X_2 &=&{1\over 4}
( Q^{\dagger} Q~ +~ L^{\dagger}L )^2~ L^{\dagger}L \nonumber \\
\label{projection2}
\sqrt{X^{\dagger}_3 ~ X_3}&=&{1\over 4} ~( Q^{\dagger} Q )^2 ~-~
{1\over 4}~( L^{\dagger} L )^2 \;
\end{eqnarray}
along the flat directions. Equations (\ref{projection1}),
(\ref{projection2}) hold for the scalar components of the superfields.
The supersymmetry of the low-energy theory lifts them to superfield
equations along the flat directions.

Using the notation of ref. \cite{ADS2},
\begin{eqnarray}
A &=& {1\over 2}\;(\; X^{\dagger}_1 ~X_1 ~+~
X^{\dagger}_2~ X_2\; )\nonumber ~,\\
\label{AB}
B &= &{1\over 3}\sqrt{X^{\dagger}_3 ~X_3} ~,
\end{eqnarray}
the K\"ahler potential (\ref{kahler}), projected onto the flat directions
becomes
\begin{eqnarray}
K ~ \bigg\vert_{\rm flat}~&=& ~\left(~Q^{\dagger}~Q
{}~+ ~\bar{Q}~\bar{Q}^{\dagger}~ +~ L^{\dagger}~L~\right)~ \bigg\vert_{\rm
flat} ~\equiv~ K_{\rm light}~(X_i^{\dagger}, X_i)~=\nonumber \\
\label{sigmak}
& = &24\;{ A ~+ ~B ~x \over x^2} \;,
\end{eqnarray}
where
\beq
\label{eks}
\; x ~\equiv~\left( Q^\dagger Q+L^\dagger L \right)~
\bigg\vert_{\rm flat}~=~4~\sqrt{B}\; {\rm cos} \left(~{1\over 3}
\; {\rm Arccos} ~ {~A~ \over B^{3/2}}~\right) \; .
\eeq
Note that equations (\ref{projection2}), used to determine $x$ as a
function of the light superfields, have several solutions. Equation
(\ref{eks}) is the only one that leads to a positive definite K\"ahler
metric at the minimum (\ref{sigmamin}).

The low-energy theory is therefore described by the sigma model with K\"ahler
potential $K_{\rm light}$ (\ref{AB}), (\ref{sigmak}), (\ref{eks}) and
superpotential
(\ref{superpotential})
\begin{equation}
\label{sigmaw}
W_{\rm light} \; = \;\lambda\; X_1 + \; 2\; {\Lambda_3^7 \over X_3}\; .
\end{equation}
In the limit (\ref{lambda}), (\ref{sigmaw}) can be treated as
a perturbation on the theory without a superpotential.

To find the ground state of the model, we must minimize the
scalar potential of the sigma model (\ref{sigmak}), (\ref{sigmaw}).
Because this potential is so complicated, we found it
convenient to minimize the potential of the full theory. In the limit
(\ref{lambda}), it suffices to minimize $V_F$ along the minima of $V_D$.
Up to local symmetries, the space of minima of $V_D$ is six-dimensional.
The two global U$(1)_Y$ and U$(1)_R$ symmetries
reduce this number to four. Numerically minimizing $V_F$ with respect
to these four parameters, one finds the minimum of the potential for the
full theory.

At the minimum, the corresponding values of the composite light fields
are
\begin{eqnarray}
X_1 &= & 0.50\; {\Lambda_3^3 \over \lambda^{3/7}} ,\nonumber \\
\label{sigmamin}
X_2 &=&0 ,\\
X_3 &= & 2.58\; {\Lambda_3^4 \over \lambda^{4/7}} .\nonumber
\end{eqnarray}
We have checked that these values also minimize the scalar potential
of the low-energy sigma model (\ref{sigmak}), (\ref{sigmaw}).

The sigma-model approach is especially useful for finding
the low-energy spectrum. The vacuum energy density is
\beq
\label{vacenergy}
M_S^4 ~=~ 3.59\;\lambda^{10/7}\; \Lambda_3^4 ~.
\eeq
The scalar mass matrix is given by\footnote{To find the masses,
one must properly normalize the kinetic terms.}
\beq
\label{VVVV}
m^2_{ab} ~=~ \langle V_{ab} \rangle
\eeq
where $V = W_i \, K^{-1}_{ij*}\, W_{j*}$ and $a,b = 1,...,6$
label the six light real fields.  It gives
three real scalar fields of masses $3.88, 2.83\;
{\rm and} \; 2.04$ (in units of $\lambda^{6/7} ~\Lambda_3$), a complex
scalar of mass  $1.35$ (in the same units), and a massless $R$ axion.
The fermion mass matrix is \cite{WB}
\beq
\label{fermionmasssigma}
m_{ij} ~=~ \langle W_{ij}~- ~K_{k\ell *}^{-1}~K_{ij\ell *}~ W_k
\rangle ~,
\eeq
where $i,j = 1,...,3$ label the three light fermions.
The fermion spectrum consists of a massless goldstino,
a massless fermion of unit hypercharge, and a fermion of mass
$3.19 \; \lambda^{6/7}\; \Lambda_3 $.

We have also performed an expansion of the full theory around the minimum,
along the lines of ref. \cite{HLW}. We integrated out
the heavy fields by substituting the solutions to their equations of motion
into the potential $V_D + V_F$. This gave a potential for the light fields,
which we minimized in an expansion in $ \lambda^2/g_{2}^2$ and
$ \lambda^2/g_{3}^2$.

To leading order in $\lambda$, the light fermion mass matrix is simply
\beq
\label{fermionmassesfull}
m_{ij}~=~ \langle W_{ij} \rangle  ~,
\eeq
where $W_{ij}$ are the derivatives of the superpotential
(\ref{superpotential}), evaluated in the unperturbed minimum (\ref{sigmamin}),
and $i,j = 1,...,3$ label the light fermion fields.

The light scalar mass matrix is more complicated. To leading order
in $\lambda$, it is
\beq
\label{lightmassesfull}
m^2_{ab}~=~\langle V_{F~ab} ~-~V_{D~abA} ~V_{D~AB}^{-1}~V_{F~B}
\rangle ~,
\eeq
where $A,B = 1,...,11$ and $a,b = 1,...,6$ label the eleven
heavy and six light real scalars, respectively;
$V_{D~AB}$ is the unperturbed heavy-scalar mass matrix; and
all derivatives of $V_F$ and $V_D$
are evaluated at the unperturbed minimum (\ref{sigmamin}).
The second term in (\ref{lightmassesfull}) is induced by the order
$\lambda^2/g_{2,3}^2$ correction to the heavy-field vevs.
We found that the light particle spectrum that follows from
(\ref{fermionmassesfull}), (\ref{lightmassesfull}) is identical
to the one presented above.

\subsection{Supergravity Couplings and $R$-Axion Mass}

In this section we couple the model to supergravity, and compute the
supergravity contribution to the $R$-axion mass. We also determine
the $R$-axion couplings and calculate its decay rate into visible
particles.

The supergravity coupling is straightforward and can be done
either in the full theory or in the effective theory of the previous section.
Since we are interested in the light sector only, we will work with the
effective theory. The results, of course, are identical to those that
are obtained with the full theory.

The most important effect of the supergravity coupling is its explicit
breaking of the $R$ symmetry. In models where all scales are much
smaller than the Planck mass\footnote{Models where the cosmological
constant is canceled by nonrenormalizable $R$-symmetric terms in the
K\"ahler potential typically require Planck-scale vevs \cite{N}.},
the only way to cancel the cosmological constant is to add a constant
term (\ref{W0MS}) to the superpotential,
\beq
\label{constantW}
W_{0}~=~{1\over \sqrt{3}}~ M_S^2 M_P ~.
\eeq
This constant might arise from a distinct sector of the theory; in
this paper we assume that it exists, but we do not address its
source.\footnote{Actually, the constant can take either sign. The sign
determines the potential of the axion field. In the stable
vacuum for the axion, this is the correct sign.}
We also assume that there are no other contributions to the
constant term in the superpotential, associated with symmetry
breaking at a higher scale, because all such phase transitions
preserve supersymmetry.

In this class of models, all soft breaking terms (gravitino mass,
scalar masses and trilinear scalar terms) are induced by $W_{0}$.
One finds terms in the scalar potential that break $R$ symmetry,
\beq
\label{v1}
V_1 ~=~{1\over \sqrt{3}}~{M_S^2 \over M_P}\left(W_{i}~K_{ij*}^{-1}~ K_{j*}~
-~3~W \right)
{}~+~{\rm h.c.} ~+~ ...~,
\eeq
where $K$ and $W$ are the K\"ahler potential (\ref{sigmak}) and superpotential
(\ref{sigmaw}) of the effective theory, and the dots denote terms
suppressed by additional powers of $M_P$. The gravitino mass is
\beq
\label{m3/2}
m_{3/2}~=~{ W_{0} \over M_P^2}~=~
{1\over \sqrt{3}}~{M_S^2 \over M_P}~=~
1.09~ {\lambda^{5/7} \; \Lambda_3^2\over M_P} ~.
\eeq

The $R$-symmetry-breaking terms in the scalar potential also give mass to the
$R$ axion. To find the mass, it is easiest to realize the $R$ axion
nonlinearly in the effective theory,
\beq
\label{axion}
X_k ~=~\langle X_k \rangle ~ {\rm exp}\left( i r_k ~{a\over f_a} \right)~,
\eeq
where $X_k$ are the light fields (\ref{lightfields}), and $\langle X_k \rangle$
and $r_k$ are their vevs (\ref{sigmamin}) and $R$ charges (\ref{Rcharges}),
respectively. The axion coupling constant is
\beq
\label{faxion}
f_a ~=~ 2.18~{\Lambda_3 \over \lambda^{1/7}}~=~1.58 ~{M_S \over
\sqrt{\lambda}}~,
\eeq
while axion mass is
\beq
\label{rmass}
m_a^2~ = ~10.0 \; \lambda^{11/7}\; {\Lambda_3^3 \over M_P} ~=~
6.58 \sqrt{\lambda}~ m_{3/2}\; M_S \; .
\eeq
For $M_S ~\simeq~10^{11}$ GeV, they become
\beq
\label{MASS}
f_a ~ \simeq ~ 10^{11}~{\rm GeV},\qquad m_a ~ \simeq ~ 10^7~{\rm GeV} .
\eeq
The supergravity couplings also contribute a small correction, of order
$m_a$, to the particles of mass $M_S$. The supergravity couplings
do not give mass to the fermion of hypercharge one.

\subsection{Axion Interactions in RHS Models}

We now consider the axion interactions in renormalizable
hidden-sector models.  In such models the supersymmetry
breaking is communicated to the visible world by gravity. As
discussed above, the constant term in the superpotential induces
soft supersymmetry-breaking scalar masses and trilinear
terms in the visible sector, with mass parameters proportional
to $m_{3/2} \simeq 10^3$ GeV.

Let us assume that, in the limit $M_P\rightarrow\infty$, the visible sector
has an $R$ symmetry. Since in that limit the two sectors decouple, the model
has two independent $R$ symmetries -- one for the visible and one for the
hidden sector. The cosmological term in the superpotential induces terms
in the scalar potential
that explicitly break both $R$ symmetries, with strength $m_{3/2}$.

Now, the $R$ symmetry in the hidden sector is spontaneously broken
at the scale $M_S$, much higher than the scale of explicit breaking.
Therefore the hidden
sector has a pseudo-Goldstone $R$ axion, with mass $m_a$
(\ref{MASS}). In the visible sector, on the other hand, the scales of explicit
and spontaneous breaking are approximately the same. This implies that there
is no pseudo-Goldstone boson associated with the observable-sector $R$
symmetry.

The interactions of the $R$ axion with the visible fields are induced by
gravity
and are therefore suppressed by $M_P$. The leading-order terms in the scalar
potential that mix with the observable and hidden fields include
\beq
\label{mixingpotential}
{1\over M_P^2}\left( W_{{\rm obs}~i}~ K_{{\rm obs}~i*}~W_{\rm hid}^*
{}~+~W_{{\rm hid}~i}~ K_{{\rm hid}~i*}~W_{\rm obs}^* ~-~ 3~W_{\rm hid}^*
{}~W_{\rm obs}~+~{\rm h.c.}\right).
\eeq
Here $W_{\rm hid}$ is the hidden-sector superpotential (\ref{sigmaw}),
$K_{\rm hid}$ is the K\"ahler potential (\ref{sigmak}), and $W_{\rm obs}$ and
$K_{\rm obs}$ are the corresponding quantities for the observable sector.
After substituting the nonlinearly-realized $R$ axion (\ref{axion}), one
finds that the couplings in (\ref{mixingpotential}) induce three-body
decays of the $R$ axion into observable particles, with a suppression factor
of $(m_{3/2}/M_P)^2$.

In addition to the decay modes into visible particles, the $R$ axion in this
model can decay into pairs of gravitinos, with a helicity suppression factor
of $(m_{3/2}/m_a)^2$. If this is the only allowed two-body decay mode, an
axion will decay primarily into gravitinos.

In some models there are alternative modes of axion decay.
For example, the axion might have an anomalous coupling to two
U$(1)_Y$ gauge bosons with a strength of order $\alpha/4 \pi$.
Then the partial width into this mode is of order $(\alpha/4 \pi)^2
m_a^3/ f_a^2 \lsim 10^{-4} m_a^3/ f_a^2$.

Another possibility is for the axion to decay into light or massless
fermions. However, helicity suppression factors apply here as well. This
implies that the axion cannot decay into two massless fermions. Since
it is unlikely that models of dynamical symmetry breaking contain fermions
with mass between $m_{3/2}$ and $m_a$, the decay rate into fermions
is probably similar to that into gravitinos.

Finally, the axion might decay into light scalars. The minimum mass of
scalars in these models is of order $m_{3/2}$. If the axion decays
dominantly into these scalars, the cosmological problems discussed
in the next section are just transferred to the scalar fields.

\mysection{Cosmological Bounds}

The solution to the relic gravitino problem of supergravity
requires an inflationary period in the evolution of the universe.
In this section, we show that the standard inflationary scenario,
with Hubble constant $H_{\rm infl} \gg m_a$ during the
exponential expansion, can give rise
to a coherent background axion field that decays into gravitinos
before nucleosynthesis. If the axion has no alternative decay mode,
this leaves a large gravitino abundance that affects the light elements
abundances from standard big-bang nucleosynthesis.

In the previous section we found that in RHS models the axion
typically has a mass of order $10^7$ GeV.
For these models, we will derive a bound on
the reheat temperature of the universe, independent from that of the
thermally-produced gravitinos. For lighter gravitino masses,
this bound is competitive with the standard bound on the reheat
temperature, while for heavy gravitino masses, the bound is stronger.
In this section, we present the bound.

According to the standard inflationary scenario, after the end
of the exponential expansion, the $R$ axion field starts oscillating
at a time $t_{\rm osc}$ \cite{CFKRR}, when the Hubble constant
becomes comparable to its mass
\beq
\label{tosc}
 t_{\rm osc}^{-1}~\simeq~H~\simeq~m_a~\simeq~10^7~{\rm GeV}~.
\eeq
The amplitude of the oscillations is given by
\beq
\label{fosc}
f~\simeq~f_a~\simeq~10^{11}~{\rm GeV}~.
\eeq
The number density of $R$ axions in this coherently oscillating wave is
$n_a \simeq f^2 ~m_a$. Hence, at the time of reheating $t_{\rm rh}$, the ratio
of the axion number density to the entropy density is
\beq
\label{novers1}
{ n_a \over s} ~\simeq~ {2.5~f^2 ~m_a \over ~g~T_{\rm rh}^3}~
\left( R_{\rm osc}\over R_{\rm rh}\right)^3 ~,
\eeq
where $T_{\rm rh}$ is the reheat temperature, $R_{\rm osc}$
and $R_{\rm rh} $ are the scale factors at $t_{\rm osc}$ and
$t_{\rm rh}$ respectively, and $g \simeq 300$ is the number of
effectively massless degrees of freedom.
The energy density before reheating is dominated by coherent oscillations
of the inflaton field \cite{KT}.  Therefore the scale factor $R~\sim~t^{2/3}$,
which implies
\beq
\label{novers2}
{ n_a \over s} ~\simeq~ {2.5~f^2~m_a \over ~g~ T_{\rm rh}^3}~
\left( t_{\rm osc}\over t_{\rm rh}\right)^2 ~.
\eeq
Substituting
\beq
\label{treh}
t_{\rm rh}^{-1}~\simeq~H_{\rm rh}~\simeq~1.7~\sqrt{g}~
{T_{\rm rh}^2 \over M_P}
\eeq
into (\ref{novers2}), we find the axion number-to-entropy ratio after
reheating,
\beq
\label{novers3}
{ n_a \over s} ~\simeq~ 7.0~ {f^2~ T_{\rm rh}\over m_a~M_P^2}~.
\eeq

If the dominant decay mode of the axion is to gravitinos, as in the
3-2 model, its decay rate is
\beq
\label{gammaff}
\Gamma_{gg}~\simeq~ {m_a\over 8\pi}~\left({m_{3/2}\over f_a}\right)^2 ~
\simeq ~4 \cdot 10^{-11}~{\rm GeV}.
\eeq
This implies that the axions will be rapidly converted into gravitinos
before nucleosynthesis, at a temperature $T_D \simeq 5$ TeV. The abundance
and lifetime of the gravitinos are constrained by the successful
predictions for the light-element abundance from standard big-bang
nucleosynthesis.

The exact bound on the gravitino abundance is complicated, since
both the abundance and the lifetime depend on the gravitino mass.
It also depends on the
square of the initial amplitude of the oscillating axion field. Since
our assumption for $f$ was probably low, the bound might actually be
stronger (for example, in the 3-2 model, smaller $\lambda$ means larger
$f$). This bound also assumes that the axion mass is large, and
it too will be smaller in models with smaller energy at the minimum
(e.g., small $\lambda$ in the 3-2 model).

For the purpose of comparing with the bound from ref. \cite{EGN}, we
can use (\ref{novers3}) to find the ratio of the mass density of
gravitinos to the photon number density that would hold today if the
gravitinos were stable:
\begin{eqnarray}
\label{abundance}
X&\equiv&m_{3/2}~{ n_{3/2} \over n_\gamma} ~ \\
&\simeq & 1 \cdot 10^{-8}~{\rm GeV}~B(a\rightarrow gg)~\left({10^7 {\rm GeV}
\over m_a}\right)
\left({ m_{3/2}\over 10^{3}~{\rm GeV}}\right)~
\left({T_{\rm rh}\over 10^{10}~{\rm GeV}}\right)~
\left({ f \over 10^{11}~{\rm GeV}}\right)^2~ ,\nonumber
\end{eqnarray}
where $B$ is the branching ratio for the decay of the axion into gravitino
pairs.  In what follows, we take
$m_a = 10^7$ GeV and $f = 10^{11}$ GeV.

The physics behind the bound differs depending on whether the gravitino
lifetime is greater or less than about $10^4$ sec.  As an
example of each case, we will consider gravitinos of mass 300 GeV and 3
TeV.  The gravitino lifetime corresponding to a decay to a single massless
gauge boson and gaugino is
\beq
\label{width}
\tau_{3/2 } ~\simeq ~ 4 \cdot 10^{5}
{\rm sec} ~ \left({10^3~{\rm GeV}\over
m_{3/2}}\right)^3~.
\eeq

For gravitino lifetimes longer than about $10^4$ sec, the strongest bounds
on the gravitino density come from photodissociation of light elements by
the electromagnetic showers produced by gravitino decays \cite{EGN,KM}.
As an example of this case, we consider $m_{3/2}= 300$ GeV.  Then,
if the gravitino decays only to photon-photino
pairs, we find a lifetime of $1.4 \cdot 10^7$ sec from (\ref{width}).
For this lifetime, fig. 3 of ref. \cite{EGN} yields the following upper bound
on $X B_{\gamma}$,
\beq
\label{boundlightall}
X B_{\gamma} ~\lsim~ 3 \cdot 10^{-12} ~{\rm GeV} ~,
\eeq
where $B_{\gamma} = 1$ is the branching ratio for decay
into photon-photino pairs.  The gravitino abundance from
axion decays (\ref{abundance}) is
\beq
\label{calclight}
X ~\simeq~ 3 \cdot 10^{-9}~{\rm GeV}~{T_{\rm rh}\over 10^{10} {\rm GeV}}.
\eeq
Comparing (\ref{boundlightall}) and (\ref{calclight}) we see that the
axion-produced gravitinos will induce dissociation of the light elements
unless the reheat temperature is bounded by
\beq
\label{trhlightall}
T_{\rm rh}~\lsim~1\cdot 10^{7}~{\rm GeV} .
\eeq

However, axions are not the only source of gravitinos.
Gravitinos can also be produced by thermal scattering
after reheating. For the case of a $300$ GeV gravitino,
the corresponding bound on the reheat temperature is $T_{\rm rh} \lsim
1 \cdot 10^{7}~{\rm GeV} $ \cite{KM}. This is comparable with our bound
(\ref{trhlightall}) from axion production of gravitinos in RHS models.

If the gravitinos also have a
direct hadronic decay channel into gluon-gluino pairs,
so $B_{\gamma} = 1/9$,
their lifetime is $1.6 \cdot 10^6$ sec, and the corresponding bound on the
reheat temperature becomes $T_{\rm rh} \lsim 6 \cdot 10^{9}~{\rm GeV}$. This
bound
is
weaker than the bound from thermal production.

For gravitino with lifetime less than about $10^4$ sec, photodissociation of
light elements does not occur and
nucleosynthesis is not affected by electromagnetic showers \cite{EGN,RS}.
However, gravitino-induced hadronic showers will
cause neutron-proton conversions that
change
the ratio of neutron-to-proton density
\cite{RS}. This change affects nucleosynthesis and constrains
the abundance of gravitinos.

As an example of the short-lived case, we consider a gravitino of mass $3$ TeV.
If the only kinematically-allowed decay of the $3$ TeV gravitino is to a
photon-photino pair, hadronic showers can still be induced by quark-antiquark
production from the virtual photon.  The gravitino lifetime is $1.5 \cdot
10^4$ sec, and the hadronic branching ratio is
estimated to be $B_h \simeq 1\% $ \cite{RS}.  Assuming
gravitino decay into two
jets of energy $m_{3/2}/3 \simeq 1 $ TeV, we find the bound
\beq
\label{boundheavynotall}
{ n_{3/2} \over s}~\lsim~ 1\cdot 10^{-12},~ 4 \cdot 10^{-14}
\eeq
from fig. 4 of ref. \cite{RS}.  The
first and second numbers correspond to baryon-to-photon number
ratio $N_B/N_{\gamma} \simeq 10^{-9},~3\cdot 10^{-10}$, respectively.
The gravitino abundance from axion decay (\ref{novers3}) is
\beq
\label{calculheavyall}
{ n_{3/2} \over s}~\simeq~1 \cdot 10^{-12} ~
{T_{\rm rh} \over 10^{10} {\rm GeV}}.
\eeq
Comparing with (\ref{boundheavynotall}),
we find the following bound on the reheat temperature
\begin{eqnarray}
\label{trhheavynotall}
T_{\rm rh}~\lsim~ 1 \cdot 10^{10} ,~ 3 \cdot 10^{7}~{\rm GeV}~,
\end{eqnarray}
where again the two numbers correspond to $N_B/N_{\gamma} \simeq 10^{-9},~3
\cdot 10^{-10}$.
For such a heavy gravitino, the bound on $T_{\rm rh}$ from ref. \cite{KM}
comes from the present mass density of photinos. It is weaker, of order
$10^{11}-10^{12}$ GeV.

If both the hadronic and electromagnetic decay channels are open,
the gravitino lifetime is $1.6 \cdot 10^3$ sec. For a hadronic branching ratio
$B_h = 8/9$, and decay to
two jets of energy $1.5$ TeV, the bound becomes \cite{RS}
\beq
\label{boundheavyall}
{ n_{3/2} \over s}~\lsim~ 1 \cdot 10^{-14} ,~ 3 \cdot 10^{-16}~,
\eeq
where again $N_B/N_{\gamma} \simeq 10^{-9}, ~3\cdot 10^{-10}$.
Comparing (\ref{boundheavyall}) and (\ref{calculheavyall}), we conclude that
the hadronic showers from the gravitino decay will affect nucleosynthesis
unless the reheat temperature is bounded by
\beq
\label{trhheavyall}
T_{\rm rh}~\lsim~ 1 \cdot 10^{8} ,~ 3 \cdot 10^{6}~{\rm GeV}~.
\eeq
We see that for a heavier gravitino, the bounds on the reheat temperature
from axion production can be stronger than the bounds from thermal production
\cite{KM}.

Of course, the most important assumption was that the axion decay to gravitinos
was substantial. As discussed in the previous section, this is very
model-dependent. However, for any model, we expect either a branching ratio
of at least $10^4 (m_{3/2}/m_a)^2$ into gravitinos, or a large decay rate
into some other cosmologically dangerous species.
It is interesting nonetheless
that the background axion density leads to a potentially dangerous
gravitino background in certain models of dynamical supersymmetry
breaking.

\mysection{Conclusions}

It is quite difficult to construct models of dynamical supersymmetry
breaking. In this paper we have shown that the existence of an axion
does not further constrain these models. Our point is that the
cosmological constant can (and should) be cancelled by adding a constant
term to the superpotential. This constant explicitly breaks any continuous
$R$ symmetry, and gives mass to the $R$ axion.

We have found that in visible-sector models with supersymmetry breaking
scale greater than $10^5$ GeV,
the axion is sufficiently heavy to evade astrophysical constraints.
In nonrenormalizable hidden-sector
models, the axion mass is of order the electroweak scale and can lead to
cosmological difficulties of the sort already presented by other
singlet fields.
In renormalizable hidden-sector models, the axion mass is quite large, of order
$10^7$ GeV.

In an inflationary scenario, the
axion of renormalizable hidden-sector models can be a new source of gravitinos.
If the reheat temperature after inflation is too high, the large gravitino
abundance affects the successful predictions for the light elements.
Our general conclusion, however, is that the axion in such models is
cosmologically safe.

\section{Acknowledgements} We are grateful to David Kaplan, Scott
Thomas, and especially Ann Nelson for many useful conversations.

\newpage

\nc{\ib}[3]{ {\em ibid. }{\bf #1} (19#2) #3}
\nc{\np}[3]{ {\em Nucl.\ Phys. }{\bf #1} (19#2) #3}
\nc{\pl}[3]{ {\em Phys.\ Lett. }{\bf #1} (19#2) #3}
\nc{\pr}[3]{ {\em Phys.\ Rev. }{\bf #1} (19#2) #3}
\nc{\prep}[3]{ {\em Phys.\ Rep. }{\bf #1} (19#2) #3}
\nc{\prl}[3]{ {\em Phys.\ Rev.\ Lett. }{\bf #1} (19#2) #3}

\end{document}